\documentclass[preprint,preprintnumbers,amsmath,amssymb]{revtex4}

\usepackage{graphicx}
\usepackage{dcolumn}
\usepackage{bm}
\usepackage[final]{pdfpages}
\begin{document}
\title{Field-induced polarisation of Dirac valleys in bismuth}
\author{Zengwei Zhu$^{1}$, Aur\'elie Collaudin$^{1}$, Beno\^{\i}t Fauqu\'e$^{1}$, Woun Kang$^{2}$ and Kamran Behnia$^{1}$\email{kamran.behnia@espci.fr}}
\affiliation{$^{1}$LPEM (UPMC-CNRS), Ecole Sup\'erieure de Physique et de Chimie Industrielles, 75005 Paris, France \\
$^{2}$ Department of Physics, Ewha Womans University, Seoul 120-750, Korea}

\date{July 28, 2011}

\begin{abstract}

 \textbf{Electrons are offered a valley degree of freedom in presence of particular lattice structures. Manipulating valley degeneracy is the subject matter of an emerging field of investigation, mostly focused on charge transport in graphene\cite{rycerz,xiao,min,gunlycke}. In bulk bismuth, electrons are known to present a threefold valley degeneracy and a Dirac dispersion in each valley. Here we show that because of their huge in-plane mass anisotropy, a flow of Dirac electrons along the trigonal axis is extremely sensitive to the orientation of in-plane magnetic field. Thus, a rotatable magnetic field can be used as a valley valve to tune the contribution of each valley to the total conductivity. According to our measurements,  charge conductivity by carriers of a single valley can exceed four-fifth of the total conductivity in a wide range of temperature and magnetic field. At high temperature and low magnetic field, the three valleys are interchangeable and the three-fold symmetry of the underlying lattice is respected.  As the temperature lowers and/or the magnetic field increases, this symmetry is spontaneously lost. The latter may be an experimental manifestation of the recently proposed valley-nematic Fermi liquid state\cite{abanin}.}
\end{abstract}
\maketitle

The principal challenge in the field of ``valleytronics''\cite{rycerz} is to lift the valley degeneracy of electrons in a controlled way. In the case of graphene, a number of theoretical propositions to generate a valley-polarized flow of electrons\cite{rycerz,gunlycke} have emerged. They are yet to be experimentally realized. Valley degeneracy has also been explored in two-dimensional semiconductors such as AlAs heterostructures\cite{shayegan,bishop} and the surface states of silicon\cite{eng,takashina}. These systems, in contrast with the case of graphene, are host to valleys presenting a parabolic dispersion and a significant in-plane mass anisotropy. Recently, it has been proposed that in presence of such an anisotropy, Coulomb interaction energetically favors a spontaneous imbalance in the occupation of different valleys\cite{abanin}.

In this paper, we show that in the case of semi-metallic bismuth, a valley-polarized current be easily generated with a well-oriented magnetic field. The Fermi surface in bismuth includes three cigar-shaped electron valleys lying almost perpendicular to the high-symmetry axis known as the trigonal axis\cite{edelman}. The in-plane mass anisotropy of each valley exceeds 200\cite{liu}. This exceptional feature is a consequence of a Dirac dispersion\cite{wolff}, with a quadratic correction\cite{mcclure} pulling drastically down the effective mass along two out of the three orientations. At the bottom of the band, experiment yields $m_{1}=0.0011 m_{e}$, $m_{2}=0.26m_{e}$ and $m_{3}=0.0044m_{e}$\cite{edelman} in fair agreement with the tight-biding model\cite{liu}. In the last few years, the electronic properties of bismuth in presence of a strong magnetic has become a focus of attention\cite{behnia,li,alicea,sharlai,yang}. The high-field phase diagram of the Dirac electrons in the extreme quantum limit \cite{yang} presents more lines than what is the expected in the non-interacting theoretical picture\cite{alicea,sharlai}. The origin of these additional field scales is yet to be understood.

We report on a study of angular-dependent magnetoresistance on bismuth single crystals with a mobility in the range of $10^{6}-10^{7} cm^{2}V^{-1} s^{-1}$ (see the SI section for details on sample mobility), performed in a configuration designed to tune the contribution of each of the three valleys to charge transport. In a wide range of temperature and magnetic field, the magnetoconductivity of electrons can be quantitatively described as the sum of three contributions interchangeable by a 120 degrees rotation. As the field rotates, the contribution of each valley to the total conductivity can be tuned and a valley-polarized conductivity exceeding 80 percent can be achieved.  At high temperatures and low magnetic fields, this simple description based on the assumption that there is no difference between the zero-field population of the three valleys is successful.  However,  as the temperature is lowered or the magnetic field is increased, magnetoresistance does not display the threefold symmetry of the underlying lattice and the conductivity cannot be described as the sum of three equivalent, rotationally-symmetrical channels. This spontaneous loss of valley degeneracy may be a consequence of electron-electron interaction in presence of anisotropic electron dispersion as recently suggested\cite{abanin}.
\begin{figure}
\center {\includegraphics[width=7cm]{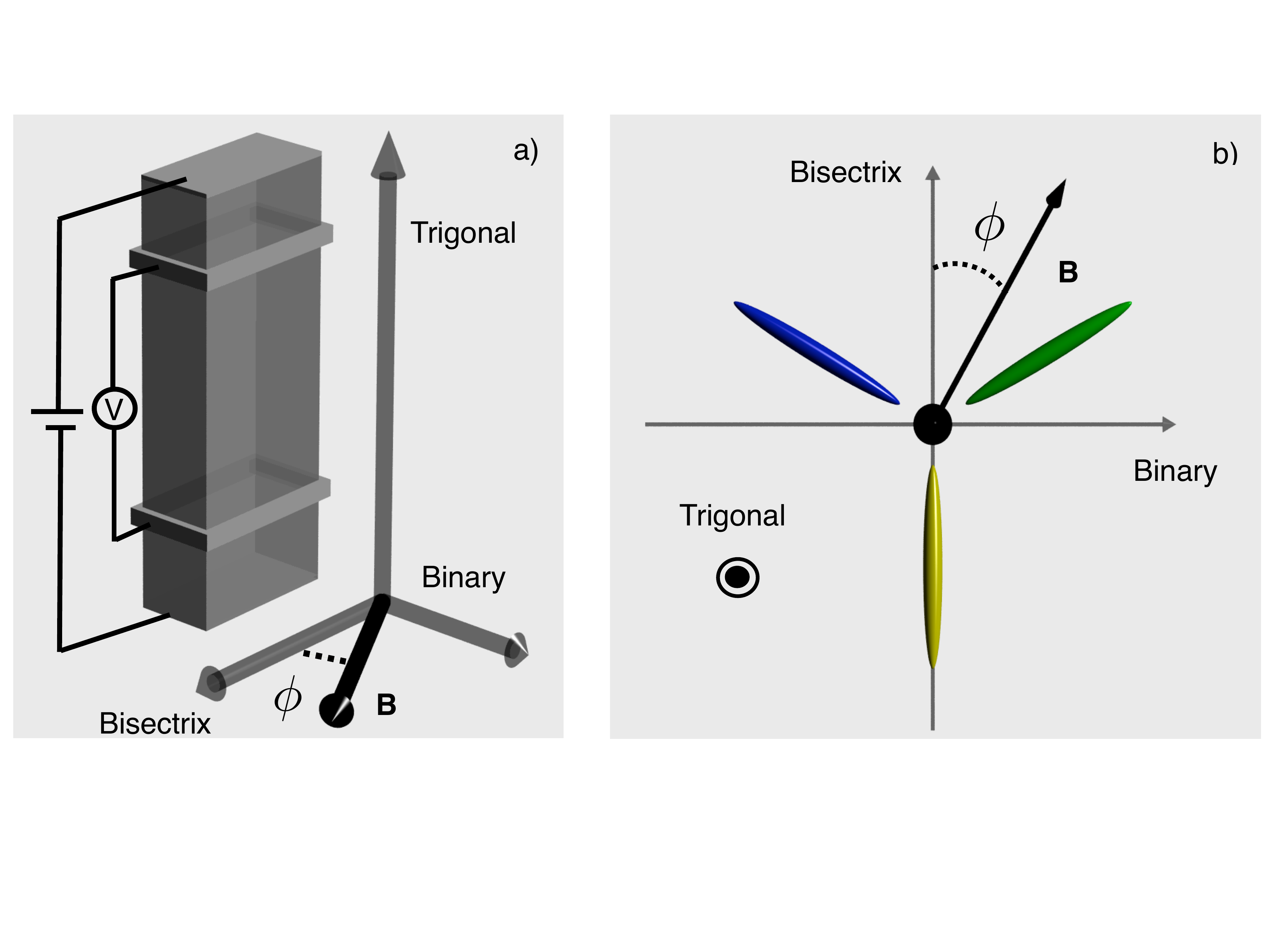}}\\
\center {\includegraphics[width=8cm]{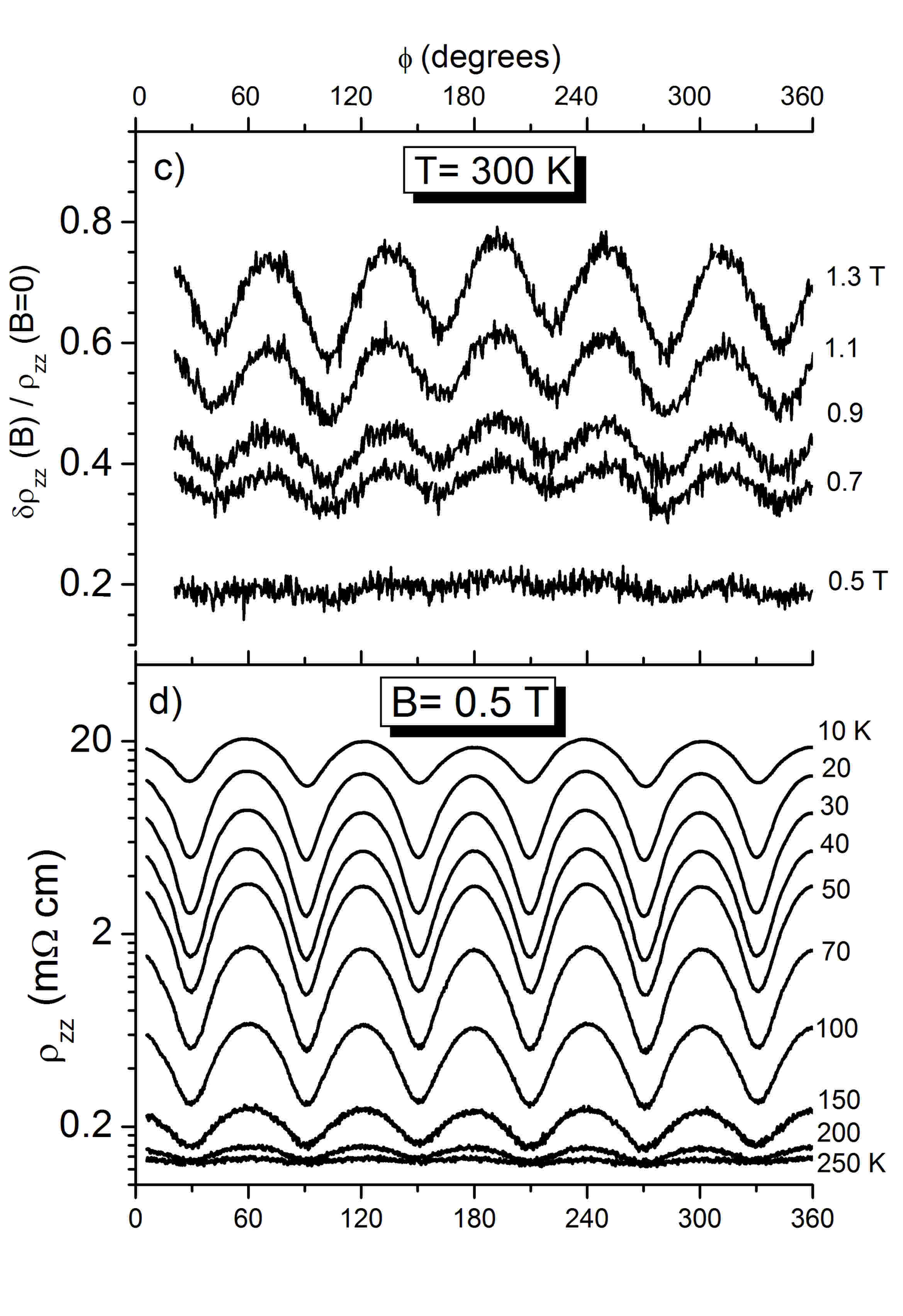}}\\
\caption{{\bf Fig. 1 The experimental configuration, the structure of the Fermi surface and angle-dependent magnetoresistance.} \textbf{a:}  Magnetoresistance of bismuth crystals was measured along the trigonal axis while the magnetic field, \textbf{B}, was rotating in the (binary, bisectrix) plane. The two vectors \textbf{B} and \textbf{J} were kept normal to each other during the rotation. \textbf{b:} The orientation of the three electron valleys with respect to the crystal axes in bismuth. \textbf{c:} Angular dependent magnetoresistance at room temperature at different magnetic fields. Oscillations are detectable when the magnetic field exceeds 0.7 T  \textbf{d:} The evolution of angle-dependent magnetoresistance at B=0.5 T with temperature. Resistivity oscillates with $\pi/3$ periodicity. It peaks each time the field is parallel or anti-parallel to one of the three bisectrix axes.}
\end{figure}
Fig. 1a shows the experimental configuration. Resistivity was measured along the trigonal axis, with a magnetic field rotating in the (binary, bisectrix) plane. In this configuration, the vector product of the two vectors, magnetic field, \textbf{B}, and current density, \textbf{J}, remains constant during the rotation. Charge carriers traveling along the trigonal axis are either holes or valley degenerate electrons. The response of these carriers to the applied magnetic field is determined by their mobility in the plane perpendicular to the orientation of the magnetic field. Because of the extreme anisotropy of the mobility tensor of each of the three Dirac valleys (Fig. 1b), angular oscillations in magnetoresistance emerge at temperatures as high as room temperature and magnetic fields as low as 0.7 T (Fig. 1c).

Fig. 1d presents a typical set of data for one of the samples at B= 0.5 T at different temperatures. Because of the lightness of electrons $\omega_{c} \tau$ is  larger than unity in the entire temperature range. Here, $\omega_{c}=\frac{eB}{m_{\perp}}$ is the cyclotron frequency, $\tau$  the scattering time, $e$ the electron charge and $m_{\perp}$ is the effective mass in the plane perpendicular to the magnetic field. The dependence of the magnetoresistance on both temperature and the orientation of the magnetic field can be qualitatively understood by considering the magnitude of $\omega_{c} \tau$. As seen in the figure, resistivity increases with decreasing temperature. This insulating-like behavior is visible for all orientations of the magnetic field\cite{du}. This is because the temperature dependence of magnetoresistance is not set by the thermal evolution of the zero-field resistivity, which  is metallic\cite{hartman} (See the SI section for details) but by the magnitude of $\omega_{c} \tau$. Since the scattering time increases with decreasing temperature, the magnetoresistance follows. The $\pi/3$ periodicity of the angular oscillations is also set by $\omega_{c} \tau$. Magnetoresistance peaks each time the magnetic field is oriented parallel or anti-parallel to a bisectrix axis. In this configuration, there is a set of carriers traveling in the plane perpendicular to the magnetic field with minimal cyclotron mass (and maximal $\omega_{c}$). These carriers are those which give rise to the largest large magnetoresistance. As the in-plane mass of the hole-like carriers is isotropic, one does not expect them to be affected by the orientation of the in-plane magnetic field.

Down to 20 K, these angular oscillations present an equal amplitude within our experimental uncertainty. Below this temperature, an inequality in the amplitude of angular oscillations grows steadily as the temperature decreases. We first focus on the regime where the threefold symmetry is preserved. Fig. 2a presents a polar plot of the resistivity data at a fixed temperature and magnetic field  (T= 40 K and B= 0.5 T). Such a presentation was first used by Mase and co-workers\cite{mase} and is particularly instructive to compare the rotational symmetries of the magnetoresistance and the underlying lattice.

In a multi-valley system in presence of a magnetic field, the electric conductivity, $\sigma$, is expected to be the sum of the contributions by individual valleys\cite{ziman}. In our particular case, one can write:

\begin{equation}\label{1}
    \sigma_{zz} = \sum_{i=1-3}\sigma_{zz}^{ei}+\sigma_{zz}^{h}
\end{equation}

where the conductivity of electron pockets and  the hole pocket  are indexed by $e_{1,3}$ and $h$. The conductivity tensor, $\overline{\sigma}$, is the inverse of the resistivity tensor, $\overline{\rho}$ and, in general, the precise determination of $\sigma_{zz}$ implies knowledge of all off-diagonal resistivity components. However, in the particular case of bismuth, a compensated semi-metal, the off-diagonal components are known to be small (i.e. $\rho_{xz}\ll \rho_{zz}$ and  $\rho_{zy}\ll \rho_{zz}$ \cite{hartman}). Therefore, setting $\sigma_{zz}= \rho_{zz}^{-1}$ is an approximation which is valid at least up to a few percent.

Fig. 2b presents a polar presentation of the angular dependence of conductivity extracted from the resistivity. We found that the simplest fit to the experimental data is given by the following function with three adjustable parameters:
\begin{equation}\label{2}
    \sigma_{zz}= \sum_{i=1-3}\frac{\sigma_{bin}}{1+ r \cos^{2}(\phi+ (i-1)\frac{2\pi}{3})} + \sigma^{h}
\end{equation}

We assume an angular-independent conductivity for the hole pocket $\sigma^{h}$.  The contribution of each of the three electron pockets is described by the same angular function rotated by $2\pi/3$. The magnetoresistance of an electron ellipsoid is lowest when the field is oriented along the binary axis which lies perpendicular to the longer axis of that particular ellipsoid. In this configuration, the magnetoconducivity of the ellipsoid in question acquires its largest value, namely $\sigma_{bin}$. As the field rotates, magnetoresistance increases and the conductivity decreases until the field becomes parallel to the longer axis of the ellipsoid, which corresponds to a bisectrix orientation. In this configuration, the magnetoresistance attains its peak and the magnetoconductance its minimal value ($\sigma_{bis}=\frac{\sigma_{bin}}{1+r}$). Thus, the parameter $r$ is a measure of the anisotropy of magnetoconductivity of an ellipsoid at a given temperature and field.

In a first approximation, the magnetoresistance is set by the cyclotron mass of electrons for a particular orientation of the magnetic field. Now, the cyclotron mass along the binary axis ($m^{bin}_{c}=\sqrt{m_2 m_3}$), is 15 times larger than  along the bisectrix axis($m^{bis}_{c}=\sqrt{m_1 m_3}$). Thus, Eq. 2 assumes an anisotropy which matches the known structure of an electron ellipsoid.

Fig. 2b displays in a polar representation the best fit to the data by this equation. As seen in the figure, the fit is satisfactory and allows a quantified observation about the relative contribution of each pocket to the total conductivity.
\begin{figure}
\center {\includegraphics[width=10cm]{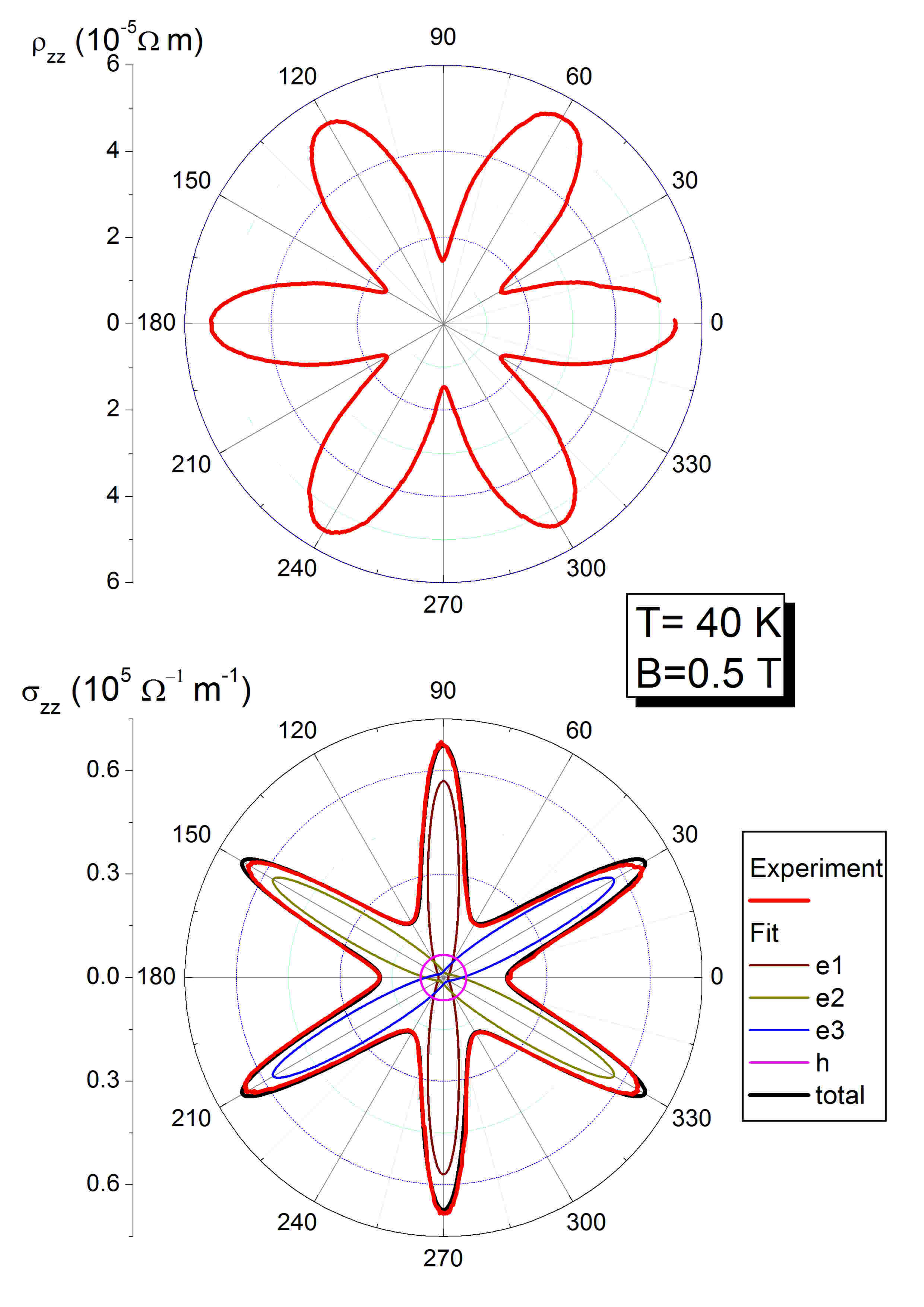}}\\
\caption{{\bf Fig. 2 The contribution of different components of the Fermi surface to the total conductivity in  polar coordinates.} \textbf{a:} Polar representation of the angular dependence of resistivity  at B=0.5 T and T= 40 K. \textbf{b:}  Polar plot of charge conductivity (red line) according to the experimental data compared to a three-parameter fit (black line) following Eq. 2. The contribution of each electron pocket (e1-e3) as well as the hole pocket (h) to the total conductivity is also plotted.}
\end{figure}
Fig. 3a shows how the magnetic field tunes the contribution of different valleys at T=40K and B=0.5 T. Since the contribution of the hole pocket is a small fraction of the total conductivity, for all orientations, charge conductivity is dominated by the flow of electrons and not holes. As the field rotates, the contribution by different valleys are modulated. When it is oriented along a binary axis, one out of the three electron pockets dominate the total conductivity. As seen in the figure, the valley-polarized conductivity exceeds 80 percent in this configuration. This is a consequence of the large magnitude of $r$, in turn a consequence of the large in-plane anisotropy of the cyclotron mass for each valley in bismuth.
\begin{figure}
\center {\includegraphics[width=8cm]{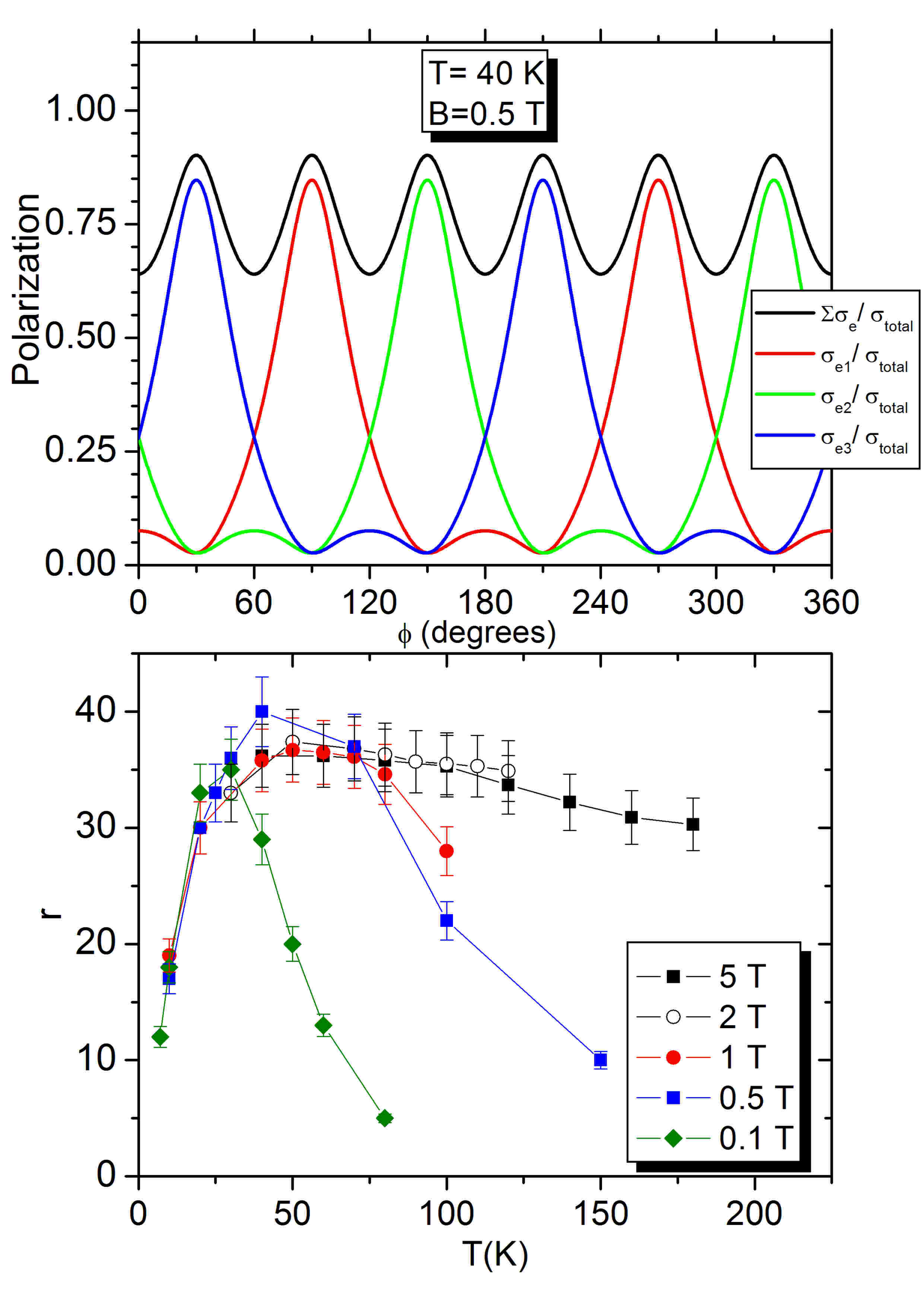}}\\
\caption{{\bf Fig. 3 Field-induced valley-polarization across a wide temperature range.} \textbf{a:}  Angular dependence of the contribution of each electron valley  as well as their sum, $\sum \sigma_{e}$, to the total conductivity,  $\sigma_{total}$. When the field is along a binary direction,  because of the large $r$, one of the valleys easily dominates the total conductivity. Moreover, the total conductivity by electrons is much larger than the contribution of the holes. \textbf{b:} The temperature dependence of $r$ for different magnetic fields. A large $r$ is attained in the intermediate temperature range. For each temperature, $r$ is obtained by finding the best fit between the data and Eq. 2. Error bars represent a window defined by the standard deviation between the fit and the data.}
\end{figure}
In a wide range of temperature and magnetic field, when the threefold symmetry of the lattice is preserved, Eq. 2 was found to be a successful fit to the data(See the supplementary Information section). This allowed us to extract the three parameters $\sigma^{h}$, $\sigma_{bin}$ and $r$  for each temperature and magnetic field. Fig. 3b plots the temperature dependence of $r$ for different magnetic fields. As seen in the figure, the peak value for $r$ (i.e. $\sim40$) is attained in the intermediate temperature range centered around 40 K. This is where the in-plane magnetic field is most efficient for generating a valley-polarized current.

Below this intermediate temperature range, $r$ decreases as the temperature lowers. This suggests a reduction in the mobility anisotropy below 30 K. Mobility is set by the ratio of the scattering time to the effective mass. The scattering time can be anisotropic in its own turn and this anisotropy can be temperature dependent. In a first approximation, its anisotropy damps the anisotropy of the effective mass. When the field is along bisectrix, the effective mass in the perpendicular plane is lighter but since the phase space for scattering is reduced, the scattering time is longer. Below 20 K, the enhancement in the anisotropy of the scattering time pulls down the anisotropy of mobility and $r$. This would indeed be the case, if  electrons residing in each valley  are scattered by those residing in another valley through Coulomb interaction as initially suggested by Hartman\cite{hartman}.

At high temperatures, $r$ also decreases. The oscillations gradually smear out as the thermal energy becomes comparable to the Fermi energy ( $\sim22 meV$) for the electrons\cite{edelman,liu}. Remarkably however, in this temperature regime, a large $r$ is gradually restored as the magnetic field increases. This suggests that orbital magnetoresistance can become an effective valley valve, even when the system exits the degenerate Fermi regime, provided that the magnetic length($\ell_{B}=\sqrt{\frac{\hbar}{eB}}$) becomes shorter than the thermal de Broglie length ($\lambda=\frac{h}{\sqrt{2\pi m^{*}k_{B}T}})$. This interesting regime has been poorly studied and deserves further exploration.

The findings reported above show that, in presence of anisotropic dispersion, the orbital magnetoresistance can tune the contribution of different valleys to the total conductivity. In multi-valley systems with modest mass anisotropy such as Si (111) or AlAs or on the (111) surface of bismuth itself\cite{ast}, this effect can be used to inject a valley-polarized flow to the two-dimensional system. However, because of the small mass anisotropy($\frac{m_{y}}{m_{x}}$ is $\approx 5$ in AlAs\cite{shayegan} and $\approx 3$ in Si(111)\cite{eng}), the effect is expected to be much less drastic than in bulk bismuth. On the other hand, due to the absence of in-plane anisotropy no such effect is expected in graphite or multi-layer graphene.

We now turn to the low-temperature high-field regime. Fig. 4b shows the thermal evolution of normalized $\rho_{zz}(\phi)$. As the temperature is lowered, the $c\overline{3}$ symmetry, the threefold symmetry of the underlying lattice, is clearly lost. We have repeated the same kind of measurements in six different crystals and with two different set-ups. As seen in the SI section, each time, the threefold symmetry, clearly present in the high-temperature data was absent in the low-temperature data, however the pattern emerging at low temperature differed from one sample to the other.

A similar evolution can be seen with increasing magnetic field at a fixed temperature as seen in Fig. 4a, which shows the data obtained on another sample. The temperature was kept at 10 K and the magnitude of the rotating magnetic field was steadily increased. As seen in the figure, the $c\overline{3}$ symmetry is preserved at B= 0.02 T and lost at B= 2 T.

We also carried out a two-axis rotation experiment on yet another sample. In this experiment, one could rotate the sample in two perpendicular planes. The relative orientation of the crystal axes and the magnetic field was scanned both in  the (binary, bisectrix) plane and in out of it. The results are presented in Fig. 5. As seen in the figure, the threefold symmetry clearly present at 30 K is lost when the single crystal is cooled down to 1.5 K. This set of data clearly rules out  the possibility that the loss of threefold symmetry is due to a residual misalignment.

Because of the lightness of electrons, the cyclotron energy in bismuth becomes large even with a modest magnetic field. The Shubnikov- de Haas effect (quantum oscillations of resistivity, which are periodic in B$^{-1}$) are visible in our data. But our findings here primarily concern the large non-oscillating background  (See the supplementary section for details). In other words, in the panels of Fig. 4, Landau quantification leads to the fine structure emerging near each lobe at low temperature and/or high field. However, the lobes themselves are not caused by the passage from one Landau level to another.

A possible explanation for the loss of threefold symmetry may be the nematic valley scenario according to which Coulomb interaction favors an unequal occupation of valleys in presence of anisotropic dispersion\cite{abanin}. A nematic liquid of electrons is formed when a metal spontaneously looses its rotational symmetry while keeping its translational invariance\cite{fradkin}. Nematic Fermi liquids have been reported in strongly correlated electron systems such as quantum Hall systems\cite{lilly}, Sr$_{3}$Ru$_{2}$O$_{7}$\cite{borzi} and more recently in URu$_{2}$Si$_{2}$\cite{okazaki}.

\begin{figure}
\center{\includegraphics[width=16cm]{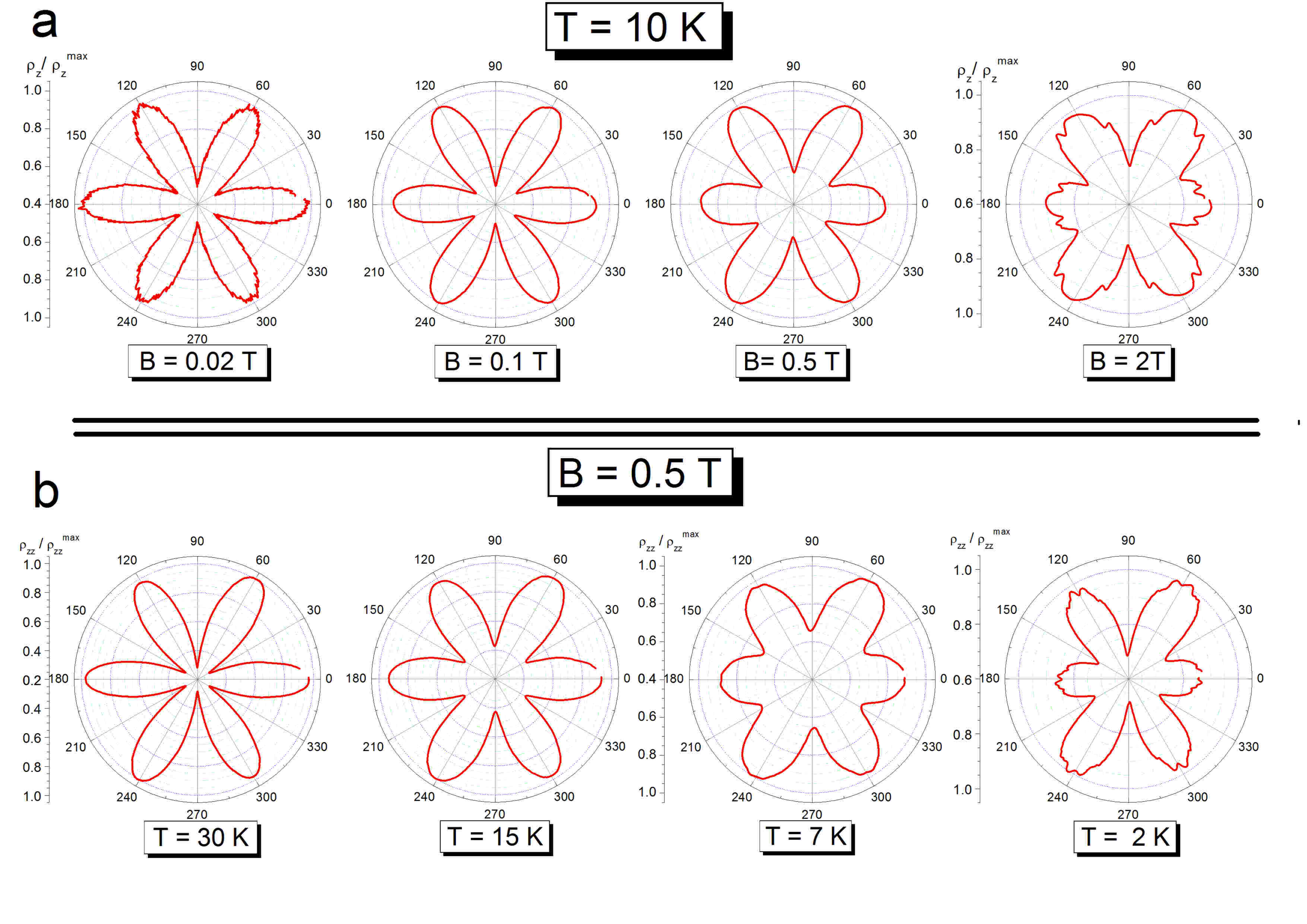}}\\
\caption{{\bf Fig. 4 Spontaneous loss of threefold symmetry by lowering the temperature or increasing the magnetic field.}\textbf{ a:} Polar representation of the angular dependence of the magnetoresistance at a fixed temperature (T= 10 K) for different fields.  At low fields the threefold symmetry of the lattice combined to the inversion symmetry generates a sixfold rotational symmetry. As the magnetic field increases, $\rho_{zz}(\phi)$ loses the three-fold rotational symmetry of the underlying lattice.  \textbf{b:} Same at a fixed field (B= 0.5 T) for different temperatures. As the temperature decreases the three-fold rotational symmetry is lost.}
\end{figure}

\begin{figure}
\center{\includegraphics[width=14cm]{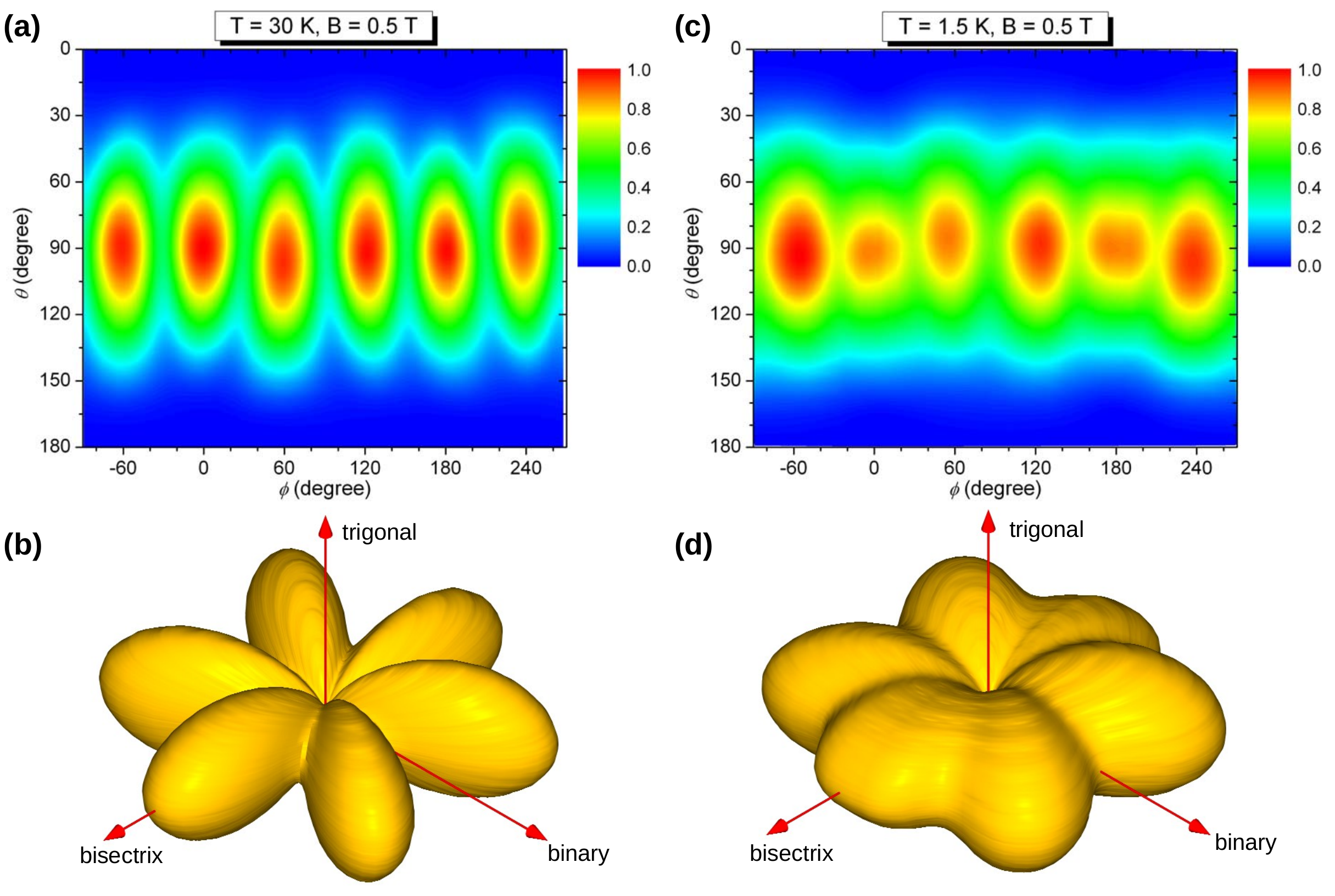}}\\
\caption{{\bf Fig. 5. Magnetoresistance data with rotation in two perpendicular planes.} \textbf{ a:} Color plot of $\frac{\rho (\theta,\phi)}{\rho_{max}}$ at T=30 K and B= 0.5 T. Here  $\theta$ is the angle between the field and the trigonal axis and $\phi$ is the angle between its projection and a bisectrix axis. \textbf{b:} Stereographic representation of the same data. \textbf{ c:} Color plot of the data for T= 1.5 K and B =0.5 T.\textbf{d:} Stereographic representation of the low-temperature data. The three valleys cease to be equivalent at low temperatures.}
\end{figure}

In our data, we do not detect any sharp anomaly in the temperature dependence at a given magnetic field and therefore, it is very hard to extract a critical temperature. Increasing the magnetic field and/or lowering the temperature regime leads to a loss of threefold symmetry in the magnetoresistance data, which is well above our experimental margin. As seen in Fig. 6, this loss of threefold symmetry becomes detectable below a characteristic temperature, which shifts upward with increasing magnetic field in agreement with what is theoretically expected.
\begin{figure}
\center{\includegraphics[width=7cm]{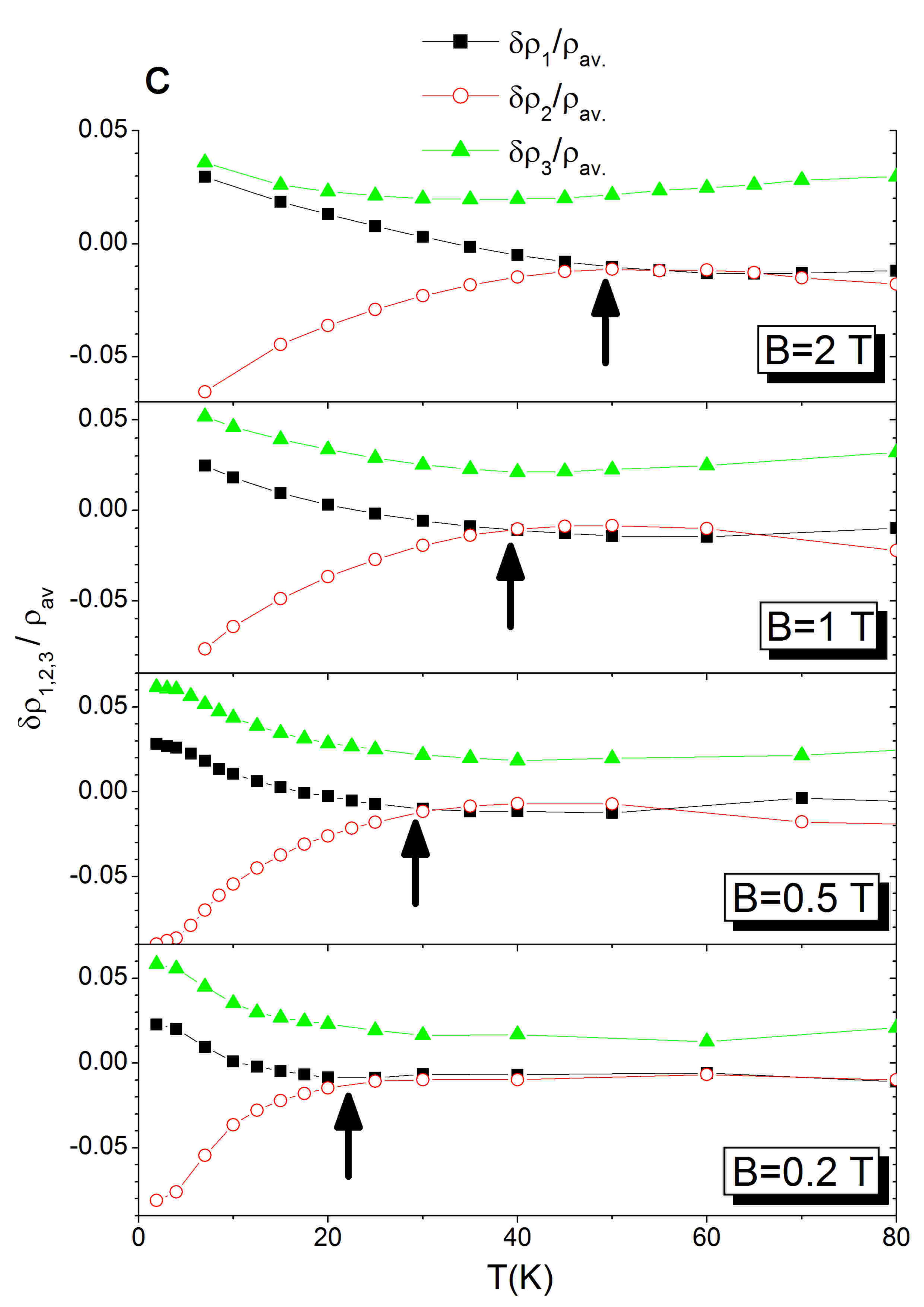}}\\
\caption{{\bf Fig. 6. The emerging  magnetoresistance anisotropy at different magnetic fields. } $\frac{\delta\rho_{i}}{\rho_{av}}$ as a function of temperature. Here $\rho_{i}$ refers to magnetoresistance resistivity for a field applied along each of the three bisectrix axes. The average resistivity is $\rho_{av}= 1/3 (\rho_{1}+\rho_{2}+\rho_{3})$ and the valley-indexed excess of resistivity is defined as $\delta\rho_{i}=\rho_{i}-\rho_{av}$. At high-temperature, because of residual misalignment between magnetic field and the trigonal plane, there is a temperature-independent difference between the three $\rho_{i}$. Below, a field-dependent threshold temperature, this difference steadily grows and saturates to a finite value due to an intrinsic loss of threefold symmetry.}
\end{figure}
We did not observe any hysteresis and each crystal reproduced the same low-temperature pattern when the measurements were repeated (See supplementary information for details). This suggests that even if Coulomb interaction  plays a role, the orientation along which the symmetry breaks is set by the crystalline imperfections of a given sample. Charge transport in bismuth is dominated by electron-electron scattering\cite{hartman} residing in extremely eccentric valleys and these features appear to be a fertile ground for the emergence of electronic liquid crystals\cite{fradkin}.

\textbf{Acknowledgements-} We thank Y. Fuseya, A. J. Millis and N. P. Ong for stimulating discussions. This work is supported  by ANR as part of DELICE and QUANTHERM projects and by a grant attributed by the \emph{Ile de France} regional council. W.K. is supported by the government of Korea through NRF of Korea Grants (2011-0000982, 0018744, 0019893). We also acknowledge STAR, a France-Korea collaboration program.

\end{document}